\documentclass[pra,amssymb,aps,showpacs,showkeys]{revtex4-2}

\usepackage{etoolbox} 
\usepackage{graphicx}
\usepackage{amsmath}
\usepackage{amssymb}
\usepackage[pdfauthor={Richard J. Mathar},colorlinks=true,pdfkeywords={phase screen, turbulence},bookmarksopen,pdfpagelayout=OneColumn]{hyperref}

\newtheorem{defn}{Definition}
\newtheorem{rem}{Remark}

\usepackage{orcidlink}

\begin{document}

\title{Simulation of Time-dependent Karhunen-Lo\`eve Phase Screens: an Ergodic Approach}

\author{Richard J. Mathar \orcidlink{0000-0001-6017-6540}}
\affiliation{Max-Planck Institute for Astronomy, K\"onigstuhl 17, 69117 Heidelberg, Germany}

\pacs{95.75.Qr, 95.75.-z, 42.25.Dd, 42.68.Bz}

\date{\today}

\begin{abstract}

Time-dependent phase screens in ground-based astronomy are
typically simulated in the so-called frozen-screen approximation
by establishing a static phase screen on a large pupil and dragging
an aperture equivalent to the size of the actual input pupil across
this oversized phase screen. The speed of this motion sweeping
through the large phase screen is equivalent to a wind speed
that changes the phase screen as a function of time.

The ergodic ansatz replaces this concept by constructing the structure function
in a three-dimensional volume---a sphere for reasons of computational efficiency---,
sampling phase screens by two-dimensional planar cuts through that
volume, and dragging them along the surface normal at some
speed which generates a video of a phase screen.

This manuscript addresses the linear algebra of populating
the three-dimensional volume with phase screens of the von-K\'arm\'an 
model of atmospheric turbulence.

\end{abstract}

\keywords{turbulence, phase screen, simulation, Taylor screen, Karhunen-Loeve}

\maketitle
\section{Phase Screens  covering two-dimensional entrance pupils}

We summarize the notation of \cite{MatharWRCM20} to define the physical
variables that establish the model of statistics of phase screens:
The phase of the electromagnetic field in the pupil plane is decomposed
into basis functions
${\mathcal K}_j$
and fluctuating expansion coefficients $a_j$
\begin{equation}
\varphi({\bf r}) = \sum_j a_j {\mathcal K}_j({\bf r}).
\label{eq.ai}
\end{equation}
The $a$ are a vector of independent scalars that are 
selected with a Gaussian random number generator for each simulation.

Integration along two parallel rays through the atmosphere 
along the lines of sight of lateral separation
$\Delta {\bf r}={\bf r}-{\bf r'}$
defines phase structure functions
that follow a power law of $\Delta r$
in the Kolmogorov limit:
\begin{equation}
\mathcal D_\varphi(\Delta r)=
\langle |\varphi({\bf r})-\varphi({\bf r'})|^2\rangle
=2c_\varphi (\Delta r/r_0)^{1+\gamma}, \quad \gamma=2/3.
\label{eq.powlaw}
\end{equation}
The scale factor is
\begin{equation}
2c_\varphi
=
2\left[\frac{8}{1+\gamma}\Gamma(\frac{2}{1+\gamma})\right]^{(1+\gamma)/2}
\approx 6.883877
\end{equation}
if $\gamma=2/3$ \cite{FriedJOSA56_1372,MatharArxiv0911}. 

Binomial expansion of the square within the expectation value relates the structure function to
the squared mean and to the covariance $C_\varphi$,
\begin{equation}
{\mathcal D}_\varphi(\Delta r) = 2 \langle \varphi \rangle ^2 - 2C_\varphi(\Delta r).
\label{eq.DofC}
\end{equation}
The Fourier representation of (\ref{eq.DofC}) is
\begin{equation}
{\mathcal D}_\varphi(\Delta r)=2\int C_\varphi(f)[ 1- \cos(2\pi {\bf f}\cdot \Delta {\bf r})]d^2f,
\end{equation}
where the imaginary
term $\sim i\sin(2\pi {\bf f}\cdot \Delta {\bf r})$ is omitted because it integrates
to zero as we assume that ${\mathcal D}_\varphi$ is isotropic, an even function of $\Delta {\bf r}$.
Circular coordinates
${\bf f}\cdot {\bf \Delta r}=f r \cos \theta$ in 
the expansion \cite[(9.1.44),(9.1.46)]{AS}
\begin{equation}
1-\cos(2\pi{\bf f}\cdot {\bf \Delta r})
=
2J_2(2\pi fr)[1+\cos(2\theta)]
+2J_4(2\pi fr)[1-\cos(4\theta)]
+\cdots,
\end{equation}
then interchange of summation and integration, plus the specification
(\ref{eq.powlaw}) lead to an inverse power law in wavenumber space \cite{NollJOSA66}
\begin{equation}
C_\varphi(f)
=
-c_\varphi
\frac{\Gamma\left(\frac{3+\gamma}{2}\right)}{\pi^{2+\gamma} \Gamma\left(-\frac{1+\gamma}{2}\right) }
\frac{1}{r_0^{1+\gamma}}f^{-3-\gamma}
\approx 0.0228955 \frac{1}{r_0^{5/3}f^{11/3}},\quad (\gamma=2/3)
.
\label{eq.Kolm}
\end{equation}
Given the covariance of the two positions in the input pupil and the
region of sampling this covariance---usually a circle of radius $\hat R$ of the primary telescope
mirror---defines
the Karhunen-Lo\`eve (KL) integral equation of the phase screens,
\begin{equation}
\iint_{r,r'\le \hat R} C_\varphi({\bf r},{\bf r'}){\mathcal K}_j({\bf r}') d^2r' = {\mathcal B}_j^2 {\mathcal K}_j({\bf r})
\label{eq.KL}
\end{equation}
with eigenvectors ${\mathcal K}_j$ and eigenvalues ${\mathcal B}_j^2$.
The variances of the coefficients $a_j$ in (\ref{eq.ai}) equal  ${\mathcal B}_j^2$
supposed we normalize each eigenvector to unity:
\begin{equation}
\iint _{r \le \hat R}|{\mathcal K}_j({\bf r})|^2d^2 r
=
\int _{0}^{\hat R}  r dr \int_{0}^{2\pi } d \phi |{\mathcal K}_j(r,\phi)|^2
=1.
\label{eq.Knorm}
\end{equation}
The index $j$
enumerates the modes.

\section{Algorithm}
A set of eigenvectors ${\mathcal K}$ established over a (circular) domain of
the input pupil produces a set of static phase screen snapshots if these
are multiplied with randomized variables $a_j$ and added up. 

\begin{figure}
\includegraphics[scale=1.5, clip, trim=1.2cm 1.5cm 1.2cm 1.2cm]{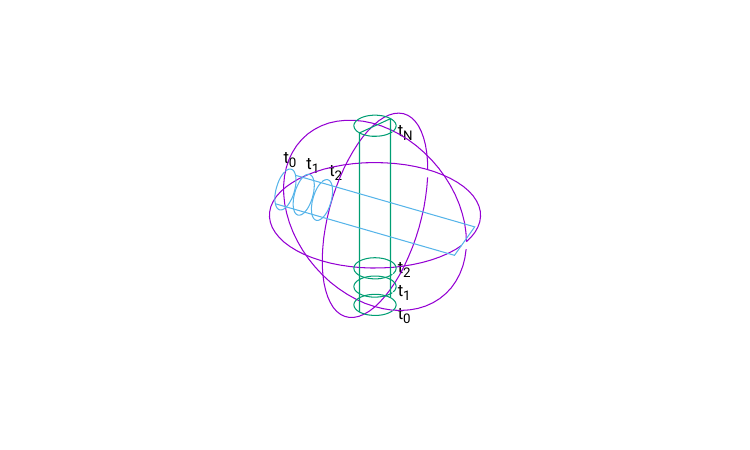}
\caption{
A 3-dimensional spherical domain outlined by three magenta circles and examples of two cylindrical
subdomains in green and in blue with different sampling
orientations. The  circular  cross sections of the 
cylinders (4 green plus 3 blue illustrated with time marks) are orthogonal to their long axes.
}
\label{fig.3d}
\end{figure}

The main idea of the implementation here is that one can look at these
snapshots as slices embedded in a 3-dimensional domain keeping exactly the
same structure function, i.e., 
the notion of 
the distances $\Delta r$ and Fourier associates $f$ defined with
the same Euclidian metrics (square roots of sums over all three Cartesian components).
Figure \ref{fig.3d} illustrates the setup: the magenta lines define
the 3-dimensional domain in which \eqref{eq.Knorm} has been solved; the four green circles
are bottom-to-top a sequence of four phase screens for a circular pupil extracted from the
interior of this domain with a time axis along the center line of the green cylinder; the
three blue circles are left-to-right another sequence of three phase screens extracted
for a time axis along the center line of the blue cylinder.

The characteristics of this methodology are:
\begin{itemize}
\item 
By construction the modes have been calculated with isotropic structure function in
all three directions; the sampling within cylindrical sub-volumes has the same statistical
features independent from axis directions or offsets.
(For phase screens of long duration one will let the cylinder axes run through the sphere center
to cover as much as possible the volume in which the modes have been set up.)
\item
The frozen screen approximation is not valid because successive circular sections
in the cylinders do not define phases which are merely shifted proportional
to the position along the axes.
\item
There is still an implicit quasi-Taylor-velocity $v$ which is needed to
define how a change in position along the cylinder axis by
a spatial $\Delta z$ translates into a time $\Delta t$
via $v=\Delta z/\Delta t$, a ratio of a coherence length and a coherence time \cite{KellererAAp461}, 
which means, how fast the circular sections are pushed
along the cylinder axes to generate a video of the phase screen.
(That velocity plays a role similar to the light velocity in special relativity
to fix a line element---although this here is simple Euclidean $2+1$ and not hyperbolic $3+1$ geometry.)
\item
The ergodic hypothesis remains enforced: statistical averages
over time for fixed points in the pupil
equal statistical averages over lateral  distances for 
fixed points in time.
\item
The algebra of the setup resembles the generation of turbulent phase distributions
in spherical domains. The structure functions of turbulent
air
are scaled with $\Delta {\bf r}^\gamma$ \cite{MatharBA17};  integration along
straight paths through 
the Earth atmosphere establishes structure functions in phase screens 
$\propto (\Delta {\bf r})^{1+\gamma}$ as in \eqref{eq.powlaw},
and we keep that $1+\gamma$ exponent to define phase screen voxels
in the model of Figure \ref{fig.3d} with 2+1 spatial+time axes.
\end{itemize}

\section{KL Evaluation} \label{sec.main}
\subsection{Primitive Basis: 3D Zernike Functions}

To keep the mathematics simple, the establishment of the KL basis function
will be carried out in a all-isotropic spherical domain, although
the number of basis functions that need to be calculated is higher
(and linear algebra numerically
more loaded) than actually needed to create merely phase screens within cylinders.

A natural choice of primitive basis functions
are the Zernike functions in $D=3$ dimensions \cite{MatharSAJ179,MatharArxiv0705a},
\begin{equation}
{\mathcal K}(r,\theta,\phi) = \sum_{n,l,m} \beta_{n,l,m} R_n^{(l)}(r/\hat R)Y_l^{(m)}(\theta,\phi)
,
\quad 0\le r\le \hat R,\quad 0\le \theta \le \pi,\quad 0\le \phi\le 2\pi
\end{equation}

\begin{defn} (3D Zernike Radial Polynomials)
Zernike Polynomials in $D$ dimensions are defined as
\begin{equation}
R_n^{(l)}(r) = \sqrt{2n+D} \sum_{s=0}^{(n-l)/2}(-1)^s \binom{(n-l)/2}{s}\binom{D/2+n-s-1}{(n-l)/2} r^{n-2s},
\label{eq.R}
\end{equation}
for azimuthal quantum numbers $n-l\equiv 0 \pmod 2$, $0\le l\le n$ and radial
Euclidean distances $0\le r\le 1$.
\end{defn}
By deliberate choice of the square root in this definition, the normalization
is
\begin{equation}
\int_0^1 r^{D-1} R_n^{(l)}(r)R_{n'}^{(l)}(r)dr =\delta_{n,n'}.
\label{eq.orthR}
\end{equation}
Only $D=3$ is relevant in this manuscript.

\begin{defn} (Associated Legendre Polynomials)
The associated Legendre functions are
\cite{MO}\cite[3.6.1(6)]{ErdelyiI}
\begin{equation}
P_l^m(x)= (-)^m(1-x^2)^{m/2}\frac{d^m}{dx^m}P_l(x),\quad m\ge 0
\label{eq.Plmdef}
\end{equation}
\end{defn}
\begin{rem}
The sign convention in \cite{JahnkeEmdeLosch}, \cite{BoschPCE25} and \cite[(4.368)]{Freeden} differs.
Some authors move the phase factor $(-)^m$ on the right hand
into the definition of $Y_l^m$ \cite{MignardAA547}.
\end{rem}
\begin{rem}
The mathematical literature extends this definition to negative $m$ by using integration
in lieu of differentiation on the right hand side (RHS) \cite[\S 4.4.2]{MO}. This is not relevant to this manuscript.
\end{rem}

\begin{defn} (Surface Spherical Harmonics)
The (complex-valued) spherical harmonics are \cite{AltmannPC53}\cite[2.1.1]{Bradley1972}
\begin{equation}
Y_l^m(\theta,\phi)=\sqrt{\frac{(2l+1)(l-|m|)!}{4\pi(l+|m|)!}}P_l^{|m|}(\cos\theta)e^{im\phi}
\label{eq.Ydef}
\end{equation}
for polar angle $0\le \theta\le \pi$, azimuth $0\le \phi\le 2\pi$ and both signs of $m$.
\end{defn}
\begin{equation}
Y_l^{m*}(\theta,\phi)=Y_l^{-m}(\theta,\phi)
.
\label{eq.Ycc}
\end{equation}
\begin{rem}\label{rem.cplx}
Some authors avoid the complex-valued $e^{im\phi}$ factor 
\cite{VityazevAL35,WieczorekGGG19}.
For simplicity of the exposition, we keep the $e^{im\phi}$ phasor;
the phase screens obtained
by superpositions ${\mathcal K}$---with
real-valued weights $a_j$ randomized for each $m$ independently---are complex-valued. If one regards
the real and imaginary parts of the phase screens as two realizations of phase screens
splitting $e^{im\phi}=\cos(m\phi)+i\sin(m\phi)$, one must multiply them
by $\surd 2$ to maintain the normalization, because the integrals over 
the squared $\sin(m\phi)$ and $\cos(m\phi)$ are only $\frac12$ (if $m\neq 0$).
\end{rem}
\begin{rem}
Schmid's equation \cite[Eq. 21]{SchmidAJP26} yields other values.
\end{rem}
These spherical harmonics are orthogonal over the surface of the unit sphere:
\begin{equation}
\int_0^\pi \sin\theta d\theta \int_0^{2\pi} d\phi Y_l^m(\theta,\phi) Y_{l'}^{m'*}(\theta,\phi)
= \delta_{l,l'}\delta_{m,m'}
,
\end{equation}
where the star means complex-conjugation and $\sin\theta$ is the Jacobian determinant
for the map from Cartesian to spherical coordinates.
\begin{rem}\label{rem.n3}
There are $2l+1$ different $m$-values for fixed $l$. There are $\lfloor (n-l)/2\rfloor$ different
$l$ values for fixed $n$, which gives 
$\sum_{l=0(1), 2\mid l-n}^n (2l+1)=(n+1)(n+2)/2$
different $(l,m)$ pairs. Including basis functions up to some $\max n$ gives
$\sum_{n=0}^{\max n}(n+1)(n+2)/2 = (1+\max n)(2+\max n)(3+\max n)/6$ distinct base functions \cite{CallahanMSMS20}.
This essentially defines the dimensions of the vector space of the linear algebra for the subsequent analysis.
\end{rem}

\subsection{Transformation to Fourier Space}
Homogeneous turbulence of the form
$C({\bf r},{\bf r'})=C({\bf r}-{\bf r'})$ implies a convolution type of
operation on the left hand side of (\ref{eq.KL}).
It is advantageous to move on to
Fourier space which transforms the convolution to a product which becomes a dyadic product
of the eigenvectors \cite{RoddierOE29}:
\begin{equation}
\int
d^3r'
\int d^3f C_\varphi(f) e^{-2\pi i {\bf f}\cdot {\bf \Delta r}}
{\mathcal K}({\bf r}')
= {\mathcal B}^2 {\mathcal K}({\bf r}).
\label{eq.KLtmp}
\end{equation}

\begin{equation}
\int
d^3r'
\int d^3f C_\varphi(f) 
e^{-2\pi i {\bf f}\cdot {\bf r}}
e^{2\pi i {\bf f}\cdot {\bf r'}}
\sum_{nlm} \beta_{n,l,m} R_n^l(r'/\hat R)Y_l^m(\hat \bf r')
= {\mathcal B}^2 
\sum_{nlm} \beta_{n,l,m} R_n^l(r/\hat R)Y_l^m(\hat \bf r])
\label{eq.KLtmp2}
\end{equation}

Normalization and phases for Fourier Transforms 
are  defined 
in this manuscript 
as
\begin{equation}
{\mathcal K}({\bf f})=\int_{}{\mathcal K}({\bf r})e^{2\pi i {\bf f}\cdot {\bf r}} d^3r
.
\end{equation}

The Rayleigh expansion of the plane wave is
\cite[(4.4)]{MatharIJQC90} \cite[(1.2)]{WenigerJMP26} \cite{BezubikJNLMP11}
$$
e^{2\pi i\bf f\cdot \bf r} 
= 4\pi \sum_{l'=0}^\infty \sum_{m'=-l'}^{l'} i^{l'} j_{l'}(2\pi fr) Y_{l'}^{m'*}({\bf f})Y_{l'}^{m'}({\bf r})
.
$$
So the Fourier Transform of the primitive bases is
(via reversal of the $m'$-summation and \eqref{eq.Ycc})
\begin{multline}
\int d^3r e^{2\pi i {\bf f}\cdot {\bf r}}
R_n^l(r/\hat R)Y_l^m({\bf r})
=4\pi \int d^3r 
\sum_{l',m'}
i^{l'} j_{l'}(2\pi fr) Y_{l'}^{m'*}({\bf f})Y_{l'}^{m'}({\bf r})
R_n^{(l)}(r/\hat R) Y_l^{(m)}({\bf r})
\\
=4\pi \int d^3r 
\sum_{l',m'}
i^{l'} j_{l'}(2\pi fr) Y_{l'}^{-m'*}({\bf f})Y_{l'}^{-m'}({\bf r})
R_n^{(l)}(r/\hat R) Y_l^{(m)}({\bf r})
\\
=4\pi \int d^3r 
\sum_{l',m'}
i^{l'} j_{l'}(2\pi fr) Y_{l'}^{-m'*}({\bf f})Y_{l'}^{m'*}({\bf r})
R_n^{(l)}(r/\hat R) Y_l^{(m)}({\bf r})
\\
=4\pi \int r^2 dr
\sum_{l',m'}
R_n^l(r/\hat R) 
i^{l'} j_{l'}(2\pi fr) Y_{l'}^{-m'*}({\bf f})
\delta_{l,l'}\delta_{m,m'}
\\
=4\pi \int r^2 dr
R_n^l(r/\hat R) 
i^{l} j_{l}(2\pi fr) Y_{l}^{-m*}({\bf f}).
\label{eq.ft}
\end{multline}
These Hankel transforms of the radial part 
(with a different normalization factor)
yield spherical Bessel Functions
\cite{LiuAC68,JanssenArxiv1510,MatharBA17},
$$
\int_0^1 R_n^l(\rho) j_l(q\rho) \rho^2 d\rho\sim (-1)^{(n-l)/2} \frac{j_{n+1}(q)}{q}
,
$$
in our case,
\begin{equation}
\int_0^{\hat R} R_n^l(r/\hat R) j_l(2\pi f r) r^2 dr
=
\hat R^3 \int_0^1 R_n^l(t) j_l(2\pi f t \hat R) t^2 dt
\\
=
(-)^{(n-l)/2} \hat R^3 \sqrt{2n+3}\frac{j_{n+1}(2\pi f\hat R)}{2\pi f\hat R}.
\end{equation}
Back in \eqref{eq.ft} the Fourier Transform of the primitive bases is
\begin{multline}
\int d^3r e^{2\pi i {\bf f}\cdot {\bf r}}
R_n^l(r/\hat R)Y_l^m({\bf r})
=
4\pi 
i^{l} 
(-)^{(n-l)/2} \hat R^3 \sqrt{2n+3}\frac{j_{n+1}(2\pi f\hat R)}{2\pi f\hat R}
Y_{l}^{-m*}({\bf f})
\\
=
4\pi 
i^{l} 
(-)^{(n-l)/2} \hat R^3 \sqrt{2n+3}\frac{j_{n+1}(2\pi f\hat R)}{2\pi f\hat R}
Y_{l}^{m}({\bf f})
.
\label{eq.ftprim}
\end{multline}

This replaces the integral over $d^3r'$ on the left hand side of \eqref{eq.KLtmp2}
\begin{multline}
\int d^3f C_\varphi(f) 
e^{-2\pi i {\bf f}\cdot {\bf r}}
\sum_{nlm} \beta_{n,l,m} 
4\pi 
i^{l} 
(-)^{(n-l)/2} \hat R^3 \sqrt{2n+3}\frac{j_{n+1}(2\pi f\hat R)}{2\pi f\hat R}
Y_{l}^{m}({\bf f})
\\
= {\mathcal B}^2 
\sum_{nlm} \beta_{n,l,m} R_n^l(r/\hat R)Y_l^m({\bf r})
.
\end{multline}
To get an eigenvalue problem for the $\beta$-numbers we project both sides
on the primitive basis functions, which means
we multiply both sides with $R_{n'}^{l'}(r/\hat R)Y_{l'}^{m'*}(r)$
and integrate both sides over $d^3r$, using again the orthogonalities of $R$ and $Y$ on the right hand
side and the Fourier transform on the left hand side:
\begin{multline}
\int d^3r
\int d^3f C_\varphi(f) 
e^{-2\pi i {\bf f}\cdot {\bf r}}
\sum_{nlm} \beta_{n,l,m} 
4\pi 
i^{l} 
(-)^{(n-l)/2} \hat R^3 \sqrt{2n+3}\frac{j_{n+1}(2\pi f\hat R)}{2\pi f\hat R}
Y_{l}^{m}({\bf f})
\\
\times
R_{n'}^{l'}(r/\hat R)Y_{l'}^{m'*}(\hat \bf r)
= {\mathcal B}^2 
\int d^3r
\sum_{nlm} \beta_{n,l,m} R_n^l(r/\hat R)Y_l^m(\hat \bf r)
R_{n'}^{l'}(r/\hat R)Y_{l'}^{m'*}(\hat \bf r)
.
\end{multline}
The $R_n^l$ on the right hand side are actually only orthogonal 
along \eqref{eq.orthR} for a common $l$, but this is fostered
by the orthogonality of the $Y_l^m$:
\begin{equation}
\int_0^{\hat R} d^2r r^2 R_{n}^{l}(r/\hat R)R_{n'}^{l}(r/\hat R)
=\hat R^3\int_0^1 d^2t t^2 R_{n}^{l}(t)R_{n'}^{l}(t)
=\hat R^3\delta_{n,n'}
.
\end{equation}

\begin{multline}
\int d^3r
\int d^3f C_\varphi(f) 
e^{-2\pi i {\bf f}\cdot {\bf r}}
\sum_{nlm} \beta_{n,l,m} 
4\pi 
i^{l} 
(-)^{(n-l)/2} \hat R^3 \sqrt{2n+3}\frac{j_{n+1}(2\pi f\hat R)}{2\pi f\hat R}
Y_{l}^{m}({\bf f})
\times
\\
R_{n'}^{l'}(r/\hat R)Y_{l'}^{m'*}(\hat \bf r)
= {\mathcal B}^2 
\hat R^3 
\sum_{nlm} \beta_{n,l,m} \delta_{n,n'}\delta_{l,l'}\delta_{m,m'}
.
\end{multline}
\begin{multline}
\leadsto \int d^3r
\int d^3f C_\varphi(f) 
e^{-2\pi i {\bf f}\cdot {\bf r}}
\sum_{nlm} \beta_{n,l,m} 
4\pi 
i^{l} 
(-)^{(n-l)/2} \sqrt{2n+3}\frac{j_{n+1}(2\pi f\hat R)}{2\pi f\hat R}
Y_{l}^{m}({\bf f})
\\
\times
R_{n'}^{l'}(r/\hat R)Y_{l'}^{m'*}(\hat \bf r)
= {\mathcal B}^2 
\beta_{n',l',m'}
.
\label{eq.eig1}
\end{multline}
The complex-conjugate of \eqref{eq.ftprim} is
$$
\int d^3r e^{-2\pi i {\bf f}\cdot {\bf r}}
R_{n'}^{l'}(r/\hat R)Y_{l'}^{m'*}({\bf r})
=
4\pi 
(-i)^{l'} 
(-)^{(n'-l')/2} \hat R^3 \sqrt{2n'+3}\frac{j_{n'+1}(2\pi f\hat R)}{2\pi f\hat R}
Y_{l'}^{m'*}({\bf f})
$$
and inserted into the left hand side of \eqref{eq.eig1}
\begin{multline}
\int d^3f C_\varphi(f) 
\sum_{nlm} \beta_{n,l,m} 
4\pi 
(-i)^{l'} 
(-)^{(n'-l')/2} \hat R^3 \sqrt{2n'+3}\frac{j_{n'+1}(2\pi f\hat R)}{2\pi f\hat R}
Y_{l'}^{m'*}({\bf f})
\\
\times 4\pi 
i^{l} 
(-)^{(n-l)/2} \sqrt{2n+3}\frac{j_{n+1}(2\pi f\hat R)}{2\pi f\hat R}
Y_{l}^{m}({\bf f})
= {\mathcal B}^2 
\beta_{n',l',m'}
.
\label{eq.d3f}
\end{multline}
This is the Fourier-domain analog of \eqref{eq.KL}, conveniently translated
to an eigenvalue problem with eigenvector components $\beta$. The numerical evaluation
of the $\int d^3f$ is described in Appendix \ref{app.df}. It is essentially
a step-by-step long write-up of the calculation in \cite{MatharBA17}.

\eqref{eq.tmp2} is
\begin{equation}
-c_\varphi
\frac{\Gamma\left(\frac{3+\gamma}{2}\right)}{\pi^{2+\gamma} \Gamma\left(-\frac{1+\gamma}{2}\right) }
\sum_{n} 
(-)^{(n'-l')/2} 
\beta_{n,l',m'} 
\sqrt{2n'+3} 
\sqrt{2n+3} 
I_{n,n'}
= \lambda ^2
(-)^{(n-l')/2} 
\beta_{n',l',m'}
\label{eq.diag}
\end{equation}
where 
\begin{equation}
\lambda^2 \equiv {\mathcal B}^2
\frac{1}{\hat R^2} (r_0/\hat R)^{1+\gamma}
\label{eq.scalB}
\end{equation}
are the eigenvalues of the matrix algebra.

The benefit of this set of variables is that a solution covers all
geometries with a common $\xi$ and a common Kolmogorov exponent $\gamma$;
from a generic table  of unitless eigenvalues $\lambda^2$ and eigenvector
components $\beta_{n,l,m}$ one can construct the physically relevant $\mathcal B^2$
by rescaling with the size parameter \eqref{eq.scalB}. The argument is
obviously the same as for construction of static/frozen phase screens 
inside circles of two  dimensions.

Because the matrix elements depend only on the radial quantum numbers $n$ and $n'$,
populating a spherical volume with KL functions---geometrically
matching the assumption of a space-time isotropic structure function---
the growth
of the number of bases in Remark \ref{rem.n3} proportional $\propto n^3$ is avoided:
the dimension of the matrix to be diagonalized grows only linearly $\propto n$.
(The total number of modes and random numbers still cover the $l$ and $m$ subspaces 
that grow $\propto n^3$.)

\begin{figure}
\includegraphics[]{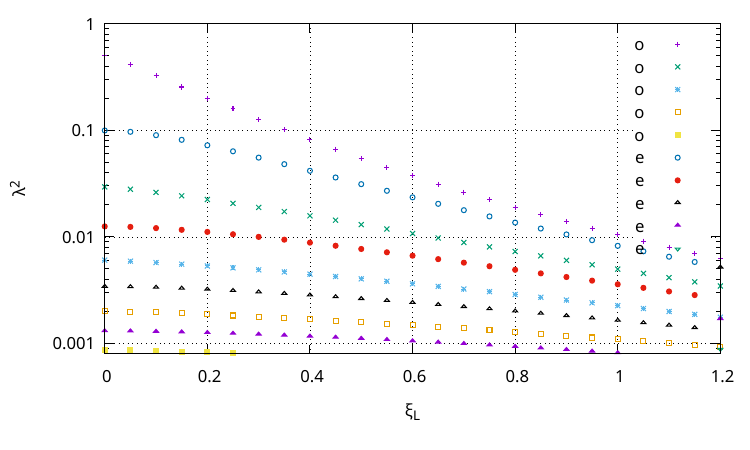}
\caption{The dominating (largest) eigenvalues $\lambda^2$ as a function of cutoff wavelength $\xi_L$.}
\label{fig.ev}
\end{figure}

An indication of the relevance of the cutoff parameter is the overview
in Figure \ref{fig.ev}. The eigenvalues $\lambda^2$ for even and odd modes
$n$ are labeled
by \texttt{e} and \texttt{o}. As already indicated in the section
of the magnitude of the matrix elements, the tip-tilt modes (of largest odd eigenvalue)
contribute with shrinking relative magnitude compared to the
other modes as $\xi_L$ increases \cite{VoitsekhoJOSAA12a}, even more so since the modes
are only added up with the weight $\lambda$, not with $\lambda^2$.
(That feature is again the same as in the static/frozen 2-dimensional phase screens.)

\begin{figure}
\includegraphics[]{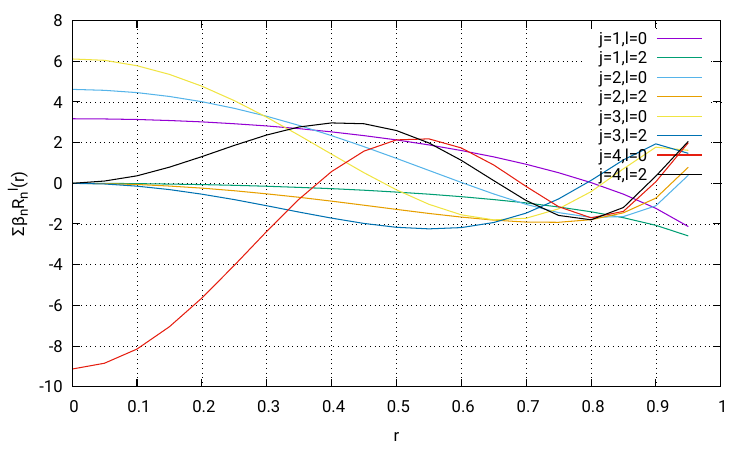}
\caption{The dominating modes for even $n$ and $n'$ (therefore even $l$) in the Kolmogorov limit.}
\label{fig.mev}
\end{figure}

\begin{figure}
\includegraphics[]{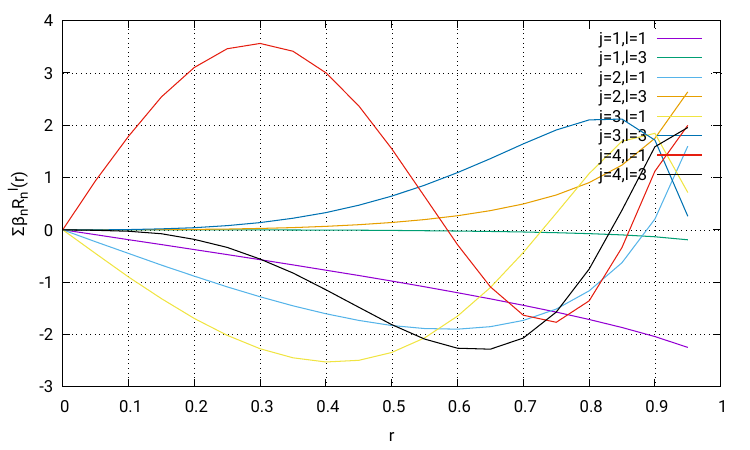}
\caption{The dominating modes for odd $n$ and $n'$ (therefore odd $l$) in the Kolmogorov limit.}
\label{fig.mod}
\end{figure}

\begin{figure}
\includegraphics[]{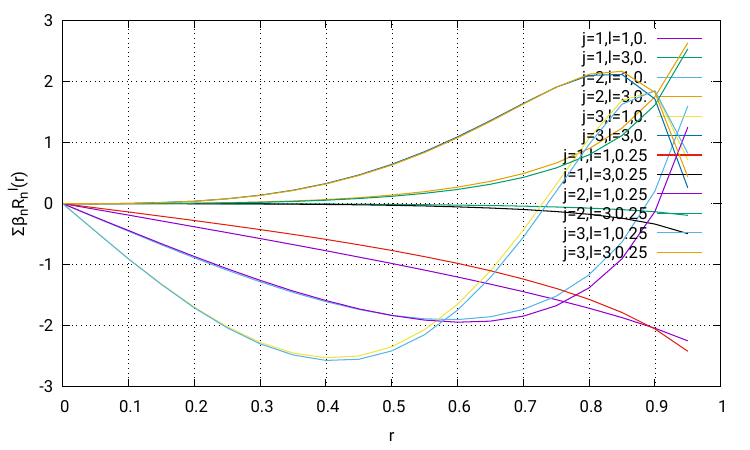}
\caption{The dominating modes for odd $n$ and $n'$ comparing 
the Kolmogorov limit $\xi_L=0$ with an (inverse) outer scale $\xi_L=0.25$}
\label{fig.modo0_5}
\end{figure}

Because the matrix on the left hand side is symmetric, the
eigenvalues are real. Because the basis functions are
complex-valued (via the dependence on the $Y_l^m$), real-valued randomized
expansion coefficients will lead to complex-valued phase screens.
If this is not desired, the real and imaginary parts can be considered
two distinct real-valued phase screens: Remark \ref{rem.cplx}.

\section{Summary} 
Phase screens defined across two-dimensional pupils can be embedded in three-dimensional
domains by keeping the structure functions such that the additional third direction
may serve as a time coordinate. Consecutive samples of two-dimensional
slices along that time coordinate
establish videos of phase screens which are self-consistent in the sense
of statistics: each slice obeys the statistics of the original
Karhunen-Lo\`eve function, and points fixed in the pupil plane have a time
statistics defined by the ergodic assumption.

We have outlined the algebra to create the Karhunen-Lo\`eve functions
for the simplest case of a three-dimensional embedding in a sphere 
and a von-K\'arm\'an clipping of the Kolmogorov power-law of the correlation of the phases.

\appendix
\section{Integrals in the wavenumber domain}\label{app.df}
Employing
the orthogonality of the $Y_l^m$ for the angular variables on the
left hand side of \eqref{eq.d3f} leaves only a radial integral and the Jacobian determinant $f^2$:
\begin{multline}
\int df f^2 C_\varphi(f) 
\sum_{nlm} \beta_{n,l,m} 
4\pi 
(-i)^{l'} 
(-)^{(n'-l')/2} \hat R^3 \sqrt{2n'+3}\frac{j_{n'+1}(2\pi f\hat R)}{2\pi f\hat R}
\\
\times 4\pi 
i^{l} 
(-)^{(n-l)/2} \sqrt{2n+3}\frac{j_{n+1}(2\pi f\hat R)}{2\pi f\hat R}
\delta_{l,l'}\delta_{m,m'}
.
\end{multline}
At this step the $\delta_{l,l'}\delta_{m,m'}$ means decoupling: the set $\beta_{n,l,m}$ is
the same for (independent of) all quantum numbers $l$ and $m$---which is the typical benefit
drawn by employing a spherically symmetric basis set 
matching the isotropic symmetry of the structure function. We rewrite $\sum_{nlm}\to \sum_n$.
Because the $\beta_{n,l,m}$ are
nonzero only if the $n-l$ are even, plus the same criterion for the $\beta_{n',l',m'}$,
the matrix elements for the vector of the $\beta$ on the left hand side are zero if $n-n'$ is odd;
effectively the problem splits into two separate sub-spaces for even and for odd $n$.
\begin{multline}
...=\int df f^2 C_\varphi(f) 
\sum_{n} \beta_{n,l',m'} 
4\pi 
(-i)^{l'} 
(-)^{(n'-l')/2} \hat R^3 \sqrt{2n'+3}\frac{j_{n'+1}(2\pi f\hat R)}{2\pi f\hat R}
\\
\times 4\pi 
i^{l'} 
(-)^{(n-l')/2} \sqrt{2n+3}\frac{j_{n+1}(2\pi f\hat R)}{2\pi f\hat R}
\\
=\hat R^3 \int df C_\varphi(f) 
\sum_{n} \beta_{n,l',m'} 
4\pi 
(-i)^{l'} 
(-)^{(n'-l')/2} \sqrt{2n'+3}\frac{j_{n'+1}(2\pi f\hat R)}{2\pi \hat R}
\\
\times 4\pi 
i^{l'} 
(-)^{(n-l')/2} \sqrt{2n+3}\frac{j_{n+1}(2\pi f\hat R)}{2\pi \hat R}
\\
=
4\hat R \int_0^\infty df C_\varphi(f) 
\sum_{n} \beta_{n,l',m'} 
(-)^{(n'-l')/2} \sqrt{2n'+3} j_{n'+1}(2\pi f\hat R)
\\
(-)^{(n-l')/2} \sqrt{2n+3} j_{n+1}(2\pi f\hat R)
\\
=4\hat R \int_0^\infty df 
(-c_\varphi)
\frac{\Gamma\left(\frac{3+\gamma}{2}\right)}{\pi^{2+\gamma} \Gamma\left(-\frac{1+\gamma}{2}\right) }
\frac{1}{r_0^{1+\gamma}f^{3+\gamma}}
\sum_{n} \beta_{n,l',m'} 
(-)^{(n'-l')/2} \sqrt{2n'+3} j_{n'+1}(2\pi f\hat R)
\\
\times (-)^{(n-l')/2} \sqrt{2n+3} j_{n+1}(2\pi f\hat R)
.
\end{multline}
For more realistic simulations a von-K\'arm\'an cut-off wavenumber $f_L$ (inverse
of the outer scale) is introduced in the denominator: 
\begin{multline}
4\hat R \int_0^\infty df 
(-c_\varphi)
\frac{\Gamma\left(\frac{3+\gamma}{2}\right)}{\pi^{2+\gamma} \Gamma\left(-\frac{1+\gamma}{2}\right) }
\frac{1}{r_0^{1+\gamma}[f^2+f_L^2]^{(3+\gamma)/2}}
\sum_{n} \beta_{n,l',m'} 
(-)^{(n'-l')/2} \sqrt{2n'+3} j_{n'+1}(2\pi f\hat R)
\\
(-)^{(n-l')/2} \sqrt{2n+3} j_{n+1}(2\pi f\hat R)
= {\mathcal B}^2 
\beta_{n',l',m'}
.
\end{multline}
We speak of $f_L\to 0$ as
the Kolmogorov limit.

A dimensionless wavenumber scale (variable substitution) is $\xi \equiv \hat R f $:
\begin{multline}
4 \int_0^\infty d\xi 
(-c_\varphi)
\frac{\Gamma\left(\frac{3+\gamma}{2}\right)}{\pi^{2+\gamma} \Gamma\left(-\frac{1+\gamma}{2}\right) }
\frac{(\hat R)^{3+\gamma}}{r_0^{1+\gamma}[\xi^2+\xi_L^2]^{(3+\gamma)/2}}
\sum_{n} \beta_{n,l',m'} 
(-)^{(n'-l')/2} \sqrt{2n'+3} j_{n'+1}(2\pi \xi )
\\
(-)^{(n-l')/2} \sqrt{2n+3} j_{n+1}(2\pi \xi)
= {\mathcal B}^2 
\beta_{n',l',m'}
.
\end{multline}

\begin{multline}
-c_\varphi
\frac{\Gamma\left(\frac{3+\gamma}{2}\right)}{\pi^{2+\gamma} \Gamma\left(-\frac{1+\gamma}{2}\right) }
\sum_{n} 
(-)^{(n'-l')/2} 
\beta_{n,l',m'} 
4 
\sqrt{2n'+3} 
\sqrt{2n+3} 
\int_0^\infty d\xi 
\frac{1}{[\xi^2+\xi_L^2]^{(3+\gamma)/2}}
\\
\times
j_{n'+1}(2\pi \xi )
j_{n+1}(2\pi \xi)
= {\mathcal B}^2 
\frac{1}{\hat R^2} (r_0/\hat R)^{1+\gamma}
(-)^{(n-l')/2} 
\beta_{n',l',m'}
.
\label{eq.tmp}
\end{multline}

\begin{rem}
The sign/phase factor $(-1)^{(n-l)/2}$ is not relevant because the $\beta$ are
eigenvector components which are only defined up to some common factor,
which is usually normalized to $\sum_{n} \beta_n^2=1$.
\end{rem}

\begin{defn} (Core Matrix Elements)
\begin{multline}
I_{n,n'}(\xi_l)\equiv 4 
\int_0^\infty d\xi 
\frac{1}{[\xi^2+\xi_L^2]^{(3+\gamma)/2}}
j_{n'+1}(2\pi \xi )
j_{n+1}(2\pi \xi)
\\
=
\int_0^\infty d\xi 
\frac{1}{\xi[\xi^2+\xi_L^2]^{(3+\gamma)/2}}
J_{n'+3/2}(2\pi \xi )
J_{n+3/2}(2\pi \xi)
.
\label{eq.intJJ}
\end{multline}
\end{defn}
In the Kolmogorov limit  this is
\cite[11.4.33]{AS}\cite{SollfreyRand}
\begin{equation}
\stackrel{\xi_L\to 0}
{\longrightarrow}
\int_0^\infty d\xi 
\frac{1}{\xi^{4+\gamma}}
J_{n'+3/2}(2\pi \xi )
J_{n+3/2}(2\pi \xi)
\\
= 
\frac{\Gamma(\frac{n+n'-\gamma}{2})\Gamma(4+\gamma)}{2\Gamma(\frac{n-n'+\gamma+5}{2})\Gamma(\frac{n'-n+\gamma+5}{2})\Gamma(\frac{n+n'+\gamma+8}{2})} \pi ^{3+\gamma}
.
\end{equation}
To simplify the notation, the two half-integer radial indices of
the Bessel Functions are rewritten as
\begin{defn} (Indices of Bessel Functions)
\begin{equation} 
Q\equiv n'+3/2;\quad T\equiv n'+3/2.
\end{equation}
\end{defn}
With this nomenclature
\begin{equation}
\int_0^\infty d\xi 
\frac{1}{\xi^{4+\gamma}}
J_Q(2\pi \xi )
J_T(2\pi \xi)
\\
= 
\frac{\Gamma(4+\gamma)\pi^{3+\gamma}
\Gamma(\frac{Q+T-3-\gamma}{2})}
{2\Gamma(\frac{Q-T+\gamma+5}{2})\Gamma(\frac{T-Q+\gamma+5}{2})\Gamma(\frac{Q+T+\gamma+5}{2})}
.
\label{eq.jj}
\end{equation}
The integral is only finite at the lower limit $\xi \to 0$ if $n+n'>\gamma$. This means
the piston mode $n=n'=0$ diverges and will be left out in the coupling matrix.

In the von-K\'arm\'an model, the integral \eqref{eq.intJJ} can be split into regions $\xi \lessgtr \xi_L$---the
approach known from near/farfield reduction of long-range potentials:
If the interval of integration is split as $\int_0^\infty d\xi = \int_0^{\xi_L}d\xi +\int_{\xi_l}^\infty d\xi $,
two terms with generalized hypergeometric functions appear:
\cite[6.541.3]{GR8}\cite{StoyanovJCAM50}\cite{SlaterHyp,WijerathnaAO62}
\begin{multline*}
I_{n,n'}(\xi_L)=
\frac{1}{2 \Gamma(\frac{3+\gamma}{2})}
(\pi\xi_L)^{T+Q} \frac{1}{\xi_L^{3+\gamma}}
\frac{\Gamma(\frac{T+Q}{2})\Gamma(\frac{3+\gamma-T-Q}{2})}{\Gamma(1+T)\Gamma(1+Q)}
\\
\times
\,_3F_4\left(\begin{array}{c}\frac{T+Q}{2},1+\frac{T+Q}{2},\frac{1+T+Q}{2}\\
1+Q,1+T,1+T+Q,\frac{T+Q-1-\gamma}{2}\end{array}\mid (2\pi \xi_L)^2\right)
\\
+\pi^{3+\gamma}
\frac{\Gamma(4+\gamma)\Gamma(\frac{T+Q-3-\gamma}{2})}
{2\Gamma(\frac{5+\gamma-T+Q}{2})\Gamma(\frac{5+\gamma-Q+T}{2})\Gamma(\frac{5+\gamma+T+Q}{2})}
\\
\times
\,_3F_4\left(\begin{array}{c}\frac{4+\gamma}{2},\frac{3+\gamma}{2},\frac{5+\gamma}{2}\\
\frac{5+\gamma-T+Q}{2},\frac{5+\gamma-Q+T}{2},\frac{5+\gamma+T+Q}{2},\frac{5+\gamma-T-Q}{2}\end{array}\mid (2\pi \xi_L)^2\right)
.
\end{multline*}

\begin{figure}
\includegraphics{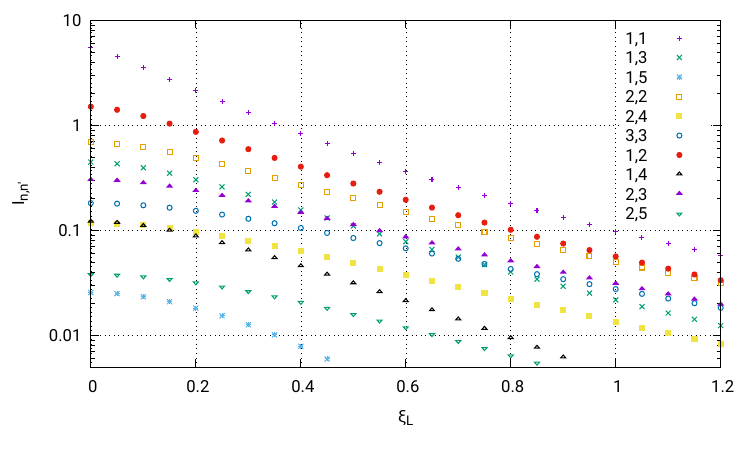}
\caption{Values $I_{n,n'}$ vor $0\le \xi_L\le 1.2$ at $\gamma=2/3$ for various pairs $(n,n')$---including 
some irrelevant ones like $(1,2)$ with odd $n+n'$.}
\label{fig.Inn}
\end{figure}

In the Kolmogorov limit $\xi_L\to 0$ both $_3F_4(;;0)=1$ in this representation; 
the first term multiplied by the factor
$\sim \xi_L^{T+Q}$ is $\to 0$ and the second term simplifies to \eqref{eq.jj}.
The general theory of hypergeometric series ensures that the power series converge
for all arguments $(2\pi \xi_L)^2$, because the lower left index (=3) is smaller than the lower right index (=4).
There are, however, large cancellation effects of the two terms if $\xi_L$ becomes large.
Characteristic values of $I_{n,n'}$ for increasing $\xi_L$ (i.e., for decreasing outer scales)
in the von K\'arm\'an model are plotted in Figure \ref{fig.Inn}.

\begin{figure}
\includegraphics{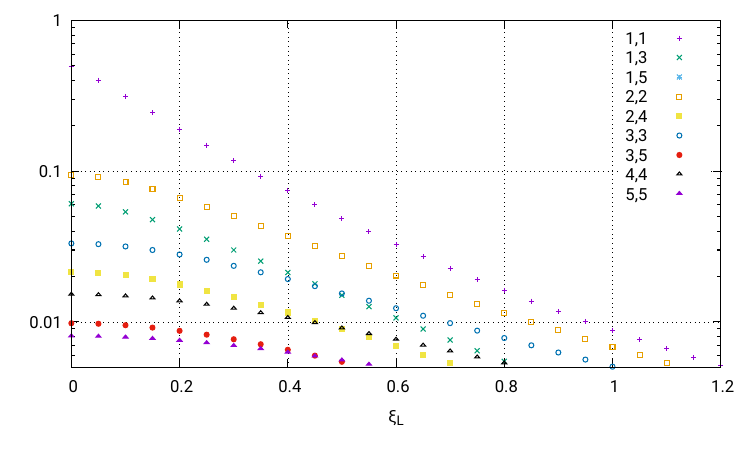}
\caption{The matrix elements in \eqref{eq.tmp2} vor $0\le \xi_L\le 1.2$ at $\gamma=2/3$ for various pairs $(n,n')$ with even $n+n'$.}
\label{fig.mat}
\end{figure}

Equation \eqref{eq.tmp} is

\begin{equation}
-c_\varphi
\frac{\Gamma\left(\frac{3+\gamma}{2}\right)}{\pi^{2+\gamma} \Gamma\left(-\frac{1+\gamma}{2}\right) }
\sum_{n} 
(-)^{(n'-l')/2} 
\beta_{n,l',m'} 
\sqrt{2n'+3} 
\sqrt{2n+3} 
I_{n,n'}
\\
= {\mathcal B}^2 
\frac{1}{\hat R^2} (r_0/\hat R)^{1+\gamma}
(-)^{(n-l')/2} 
\beta_{n',l',m'}
.
\label{eq.tmp2}
\end{equation}
Some matrix elements,
i.e., the left hand side without the sum symbols and without the $\beta$-factor,
are plotted in Figure \ref{fig.mat}.
As expected, the matrix elements of the tip-tilt-mode, $n=n'=1$, become 
less dominant over the
higher modes if $\xi_L$ increases.

\section{Support software}
\subsection{Table of Modes}
The ancillary directory \texttt{anc} contains a Maple program \texttt{KL3d.mpl}
which generates the KL modes by diagonalizing correlation matrices
in the manner of \eqref{eq.diag}. 
\begin{rem}
For large arguments $(2\pi \xi_L)^2$ Luke has provided representations 
of hypergeometric functions
in series of the inverse argument
\cite[\S 7.4.3]{Luke1969Book}. These have not been implemented here.
\end{rem}
It prints the modes in
the XML format, standard output typically redirected in the style of
\begin{verbatim}
maple -q KL3d.mpl > KL3d.xml
\end{verbatim}
The prototypical \texttt{KL3d.xml} is also reproduced in the directory. 
It has entries with \texttt{vKarman}
elements, each a mode set for a fixed specific $\xi_L$---which is the value
in the \texttt{cutoff} element. Within each of these modes the $\lambda^2$ is 
in the \texttt{eval} element, the parity of $n$ is in the \texttt{parity} element,
and the expansion coefficients
$\beta_{n,l,m}$ of this mode are listed---ordered 
by increasing $n$ of the basic radial $R_n^{m}$ polynomial---in the \texttt{coeff} element.
The modes are either even or odd; the $n$ of the basic
polynomial is added as a comment in front of each $\beta_n$.
A formal specification is in the \texttt{KL3d.dtd} file.

A rule of thumb:
If we expect outer scales at least of the order of 10 or 20 meters in ground-based
optical astronomy \cite{TokovininMNRAS378}, the von-K\'arm\'an wavenumbers $f_L$ are in the range of $0.05$ to $0.1$ 
$m^{-1}$. To apply the theory
up to telescopes of the Extremely Large Telescope (ELT) class with $\hat R\approx 20$m,
dimensionless $\xi_L=\hat Rf_L$ in the range $0$ to $2$ suffice.

The list of $\xi_L$ values in \texttt{anc/KL3d.xml} can be extracted
on the \texttt{bash} with \verb+grep cutoff KL3d.xml+.

\subsection{C++ Reader of Modes}
If the \texttt{xerces-c} library and the GNU scientific library (GSL) are make available
in the Linux system, for example with
\begin{verbatim}
zypper install xerces-c-devel gsl-devel autoconf automake
\end{verbatim}
under openSUSE, a template C++ implementation of creating modes can be compiled with
\begin{verbatim}
autoreconf -i -f -s
./configure
make
\end{verbatim}
in the \texttt{anc} subdirectory. This compiles a ELF binary $kl3d$ which
has the following modes of operation:
\begin{itemize}
\item

\texttt{kl3d -t }\textit{[-c cutoff]} \textit{[-x xmlfile]}

characterizes the mode with $\xi_L$ provided by the \textit{cutoff} number (default is 0.0)
by extracting parameters from the XML file specified by the name in the \texttt{-x} option (default 
\texttt{KL3d.xml}). \textit{Characterizes} means it prints the cutoff value, a colon and then
a short overview of the modes, one per line.
The overview shows, a blank as the spacer, the eigenvalue $\lambda^2$, the parity (0 for even, 1 for odd),
and the first few $\beta_{n,l,m}$ expansion coefficients in the Zernike basis.

This source code serves as a demonstrator how a XML ASCII file (in the current
specification) can be transformed into a set of modes represented by the \texttt{Kl3dModeSet} class.

\item

\texttt{kl3d -R }\textit{[-n n] [-l l] [-r deltar]}

tabulates values of $R_n^{(l)}(r)$ from $r=0$ up to $r=1$ in steps of \textit{deltar}.
The default values for $n$ and $l$ are 1 (the tip-tilt Zernike mode)
and 0.02 for the step in the radial direction.
\item

\texttt{kl3d -m j }\textit{[-c cutoff]} \textit{[-x xmlfile]} \textit{[-l l] [-r deltar]}

tabulates values of $\lambda_j \sum_n \beta_n R_n^{(l)}(r)$ from $r=0$ up to $r=1$ 
in steps of \textit{deltar} for mode number \textit{j}.
The enumeration for the mode numbers starts separately at 0 for 
each cutoff parameter. In the current organization of \texttt{Kl3d.xml}
the even nodes come first, the odd nodes later.
To recognize the maximum value of $j$ one may call the program with a very large $j$ and
read the error message which prints the number of modes available in the \textit{xmlfile}.

The default for \texttt{cutoff} is 0, the default for $l$ is 1, the default for \textit{deltar} is 0.02.

If the value of $l$ is larger than the support (largest) of the $n$ in the mode, all values are zero.

\end{itemize}

If \texttt{doxygen} is available, a overview of the C++ classes can be generated
in the source code directory with
\begin{verbatim}
cd anc
doxygen
firefox html/index.html &
\end{verbatim}
\section{Numerics of the Radial Polynomials}
The Zernike radial polynomials \eqref{eq.R} are essentially terminating Gaussian Hypergeometric 
series
\begin{equation}
R_n^{(l)}(r)=(-1)^{(n-l)/2} \sqrt{2n+D} \binom{(n+l+D)/2-1}{(n-l)/2} 
r^l{}_2F_1\left(\begin{array}{c}-(n-l)/2, (n+l+D)/2 \\ l+D/2\end{array}\mid r^2\right).
\end{equation}
If $l$ is kept constant and $n$ is increased in steps of two, the first upper parameter of the $_2F_1$
increases by one, the second decreases by one.
An associated contiguous relation 
\begin{equation}
(a-b)[(1+a-b)(1+b-a)(1-z) +a(a-c)+b(b-c)+c-1]F= 
\\
-a(c-b)(a-b-1)F(a^+,b^-)
+b(a-b+1)(a-c)F(a^-,b^+)
\end{equation}
can be stitched from the fundamental contiguous relations \cite{RakhaCMA61,ChoEAMJ15}
and distills the 3-term recurrence \cite{MatharArxiv0705a}
\begin{multline}
-(1+\frac{n-l}{2})(1-n-\frac{D}{2})\frac{n+l+D}{2}
\frac{R_{n+2}^{(l)}(x)}{\sqrt{2(n+2)+D}}
=
\\
\frac{n-l}{2}(1+n+\frac{D}{2})(1-\frac{n+l+D}{2})
\frac{R_{n-2}^{(l)}(x)}{\sqrt{2(n-2)+D}}
\\
+(n+\frac{D}{2})\left[(1+n+\frac{D}{2})(1-n-\frac{D}{2})(1-x^2) +\frac12(n-l)(D+n+l)+l+\frac{D}{2}-1\right]
\frac{R_n^{(l)}(x)}{\sqrt{2n+D}}
. 
\end{multline}
For $D=2$ this was published by Kintner \cite{KintnerOA23},
for $D=3$ by Deng and Gwo \cite{DengACM2020}. For fixed $l$
it produces a ladder of $R_n^{(l)}$-values starting at 
\begin{equation}
R_l^{(l)}(r)=\sqrt{2n+D}r^l;
\end{equation}
\begin{equation}
R_{l+2}^{(l)}(r)
=-\sqrt{2(l+2)+D}r^l\left[l+D/2-(l+1+D/2)r^2\right].
\end{equation}
This recurrence is useful 
as the alternating series of the monomials suffers numerically from cancellations of digits
for large $n$ \cite{HoudayerHDMIalg15}. 
The C++ implementation is the function \texttt{atNlist} in the class \texttt{Zern3dR}.

Alternatively, if one wishes to keep $n$ constant and to derive a ladder of values for $l=n, n-2, n-4\ldots$,
the upper two parameters of the hypergeometric function increase by one and the lower parameter
increases by two if $l$ increases by two.
In this case a supporting 3-term contiguous equation is
\begin{equation}
c(c^2-1)[c(c-2)-(cb-2ba-c+ca)z]F=
c^2(c^2-1)
(c-2)F(a^-,b^-,c^{--})
\\
+ab(c-2)(c-b)(c-a) z^2F(a^+,b^+,c^{++}). 
\end{equation}
The parameter set $a=-(n-l)/2$, $b=(n+l+D)/2$ and $c=l+D/2$ yields
\begin{multline}
(l+\frac{D}{2})(l+1+\frac{D}{2})(l-1+\frac{D}{2})\left[(l+\frac{D}{2})(l-2+\frac{D}{2})-\frac12(l^2+lD+Dn+D^2/2+n^2-2l-D)z\right]
\\
\times {}_2F_1\left(\begin{array}{c}-(n-l)/2,(n+l+D)/2 \\ l+D/2\end{array}\mid z\right)
\\
=
(l+\frac{D}{2})^2(l+1+\frac{D}{2})(l-1+\frac{D}{2})(l-2+\frac{D}{2})
{}_2F_1\left(\begin{array}{c}-(n-l+2)/2,(n+l-2+D)/2 \\ l-2+D/2\end{array}\mid z\right)
\\
+
(\frac{n-l}{2})^2(\frac{n+l+D}{2})^2(l-2+\frac{D}{2}) z^2
{}_2F_1\left(\begin{array}{c}-(n-l-2)/2,(n+l+2+D)/2 \\ l+2+D/2\end{array}\mid z\right)
,
\end{multline}
which gives a 3-term recurrence coupling $R_n^{(l)}$ on the left hand side
with $R_n^{(l-2)}$
and $R_n^{(l+2)}$
on the right hand side.
In $D=2$ dimensions this recurrence was proposed by Chong et al.\ \cite{ChongPR36}.

\bibliography{all}

\end{document}